# Ultimately deformed double-network gels possess positive energetic elasticity

Chika Imaoka,[1] Tatsunari Masumi,[1] Jian Ping Gong[2,3], Tsutomu Indei,[2*] Tasuku Nakajima,[2,3*]

[1]Graduate School of Life Science, Hokkaido University, Sapporo, Hokkaido, 001-0021, Japan
[2]Faculty of Advanced Life Science, Hokkaido University, Sapporo, Hokkaido, 001-0021, Japan
[3]The Institute of Chemical Reaction Design and Discovery, Hokkaido University, Sapporo, Hokkaido, 001-0020, Japan

**Abstracts**

The elasticity of rubbery polymer networks has been considered to be entropy-driven. On the other hand, studies on single polymer chain mechanics have revealed that the elasticity of ultimately stretched polymer chains is dominated by the energetic contribution mainly originating from chemical bond deformation. Here, we experimentally found that the elasticity of the double-network gel transits from the entropy-dominated one to the internal energy-driven one with its uniaxial deformation through the thermodynamic analysis. Based on this finding, we developed a simple mechanical model that takes into account the energetic contribution and found that this model approximately reproduces the temperature dependence of the stress-strain curve of the double-network gel. This study demonstrates the importance of the chemical perspective in the mechanical analysis of highly deformed rubbery polymer networks.

**Article text**

**I. Introduction**

Understanding the elasticity of polymeric systems ranging from a single polymer chain to a rubbery polymer network is a central topic of polymer physics. Thermodynamics dictates that the elastic force of a material originates from the change in its free energy. For a uniaxial deformation of a material under an isovolumetric process, the following equation is generally valid:

$$f = \left(\frac{\partial U}{\partial L}\right)_{T,V} - T\left(\frac{\partial S}{\partial L}\right)_{T,V} \tag{1}$$

where $f$, $T$, $U$, $L$, $V$, and $S$ are the elastic force, absolute temperature, internal energy, length, volume, and entropy, respectively. [1] This equation tells us the two origins of elasticity, which are the energetic contribution $\left(\frac{\partial U}{\partial L}\right)_{T,V}$ and entropic contribution $-T\left(\frac{\partial S}{\partial L}\right)_{T,V}$.

Through the establishment of classic rubber elasticity theories starting from the discovery of the Gough-Joule effect, the elasticity of rubbery networks and single polymer chains has been considered entropy-dominated. [2] The freely jointed chain (FJC) model has been established as a simple model to express the entropy-dominated elasticity of a polymer chain. [3] This model considers the contribution of the conformational entropy of a polymer chain but does not consider any energetic contributions, such as the deformation of chemical bonds in the main chain and intramolecular interactions. A polymer chain obeying the FJC model, also called an inverse-Langevin chain, exhibits a linear force-displacement relationship in the small deformation regime and extreme hardening near the stretching limit. Extending the FJC model to the three-dimensional polymer network, constitutive hyperelastic models for rubbery polymer networks, such as the Neo-Hookean model and the Arruda-Boyce model, have also been constructed. [4,5] These simple 3-D extensions also only consider the entropic contribution and ignore any energetic contribution. [6,7]

Apart from these classical theoretical treatments, experimentalists have developed a way to stretch a single polymer chain with an atomic force microscope (AFM). The experimentally obtained force-extension curves of real polymer chains are remarkably different from those of FJC model chains. Real polymer chains show milder hardening near their stretching limit than expected by the FJC model [6,7] Moreover, the shape of a force-extension curve of real polymers near the high extension limit is strongly affected by their chemical characteristics. [8–10] These results mean that the deformation of chemical bonds, which leads to positive energetic contributions, strongly affects the elasticity of an ultimately stretched polymer chain. To simply express both energetic and entropic contributions of the elasticity of a polymer chain, the modified FJC (m-FJC) model has been established, and it adequately reproduces the force-extension curves of many real polymer chains. [6,11–13] It has been pointed out that most of the elastic energy stored in the ultimately stretched single polymer chain originates from the energetic contribution. [14]

Nowadays, the remarkable energetic contribution originating from the chemical bond deformation to the elasticity of ultimately stretched polymer chains seems to be widely accepted. If so, the energetic contribution should also dominate the elasticity of an ultimately stretched rubbery polymer network. However, there has been no experimental report on a strong positive energetic contribution to the elasticity of a deformed real rubbery network. The ratio of the energetic and entropic contributions to the elasticity of real rubbers and gels has been evaluated by measuring the temperature dependence of the stress at a fixed length, and the ratio of the energetic contribution to the whole elastic stress has been reported to be negligible or only at most 30%. [15–18] Based on these reports, constitutive hyperelastic models that only consider the entropic contribution have been still used frequently. Note that the energetic elasticity of rubbery networks with significant intermolecular interactions, such as biopolymer networks, is often assumed [19]. The energy elasticity is also experimentally evident in rubbery materials showing strain-induced crystallization [20]. In this study, we do not focus on such chemistry-specific interactions but consider *ordinary* polymer networks without significant intermolecular interactions. We also note that the (positive) energetic elasticity discussed here differs from the recently reported "negative" energetic elasticity of polymer hydrogels by Sakai *et al.* [21,22]. This negative energetic elasticity of hydrogels is due to the polymer-solvent interaction and is found regardless of the degree of deformation, while the energetic elasticity discussed here is due to the deformation of chemical bonds of a highly stretched polymer.

The lack of experimental reports of the remarkable energetic elasticity of rubbery networks might be attributed to the difficulty of deforming real rubbery networks to near their ideal deformation limit. Ideally, a rubbery network is deformable to the ultimate state where (majority of) its network strands reach their stretching limit. However, a real rubbery material cannot be deformed to this ultimate state because of its defect sensitivity. [23] When a typical rubbery material, containing defects of various scales, is deformed, stress concentration on these defects induces catastrophic crack propagation and material fracture. The actual deformation limit of typical rubbery materials is thus much smaller than their ideal deformation limit, and most of their network strands except near the crack are not stretched to the ultimate regime where energy elasticity is dominant even at the time of material fracture. Therefore, the elasticity of a real rubbery material up to its experimental fracture point is dominated by the entropy elasticity and can be adequately reproduced by a hyperelastic model that only considers the entropic contribution such as the Arruda-Boyce model. [24]

On the other hand, the double-network gel (DN gel) is a rubbery material that can be stretched to the ideal deformation limit of its primary network. The DN gel consists of contrasting two interpenetrating networks: the stiff and brittle first, primary network and the soft and stretchable second, supportive network. [25,26] Since the supportive network can relieve the stress concentration on defects of the primary network, the DN gel is defect insensitive and can be deformed to the ultimate state where the network strands of its primary networks reach their stretching limit. [27,28] Such ultimate deformation (and following fracture) of the primary network leads to several unique mechanical behaviors of the DN gel, such as extreme strain hardening, [29,30] yielding, [28,31] inverse mechanical-swelling coupling, [32] and deformation-induced radical generation inside the gel. [33] Typically, stresses of a deformed DN gel are determined almost exclusively by the stiff primary network, and the influence of the soft supportive network is not so apparent. [28,34]

Here, we experimentally found that the elasticity of the highly stretched DN gel is dominated by the positive energetic contribution. This is the first experimental demonstration of the strong positive energetic elasticity on the elasticity of an ultimately deformed polymer network without strong intermolecular interactions.

## II. Separation of the entropic and energetic contributions

Following the previous reports, we adopt the thermodynamics way to separate the entropic and energetic contributions. Elasticity of materials is usually described as stress-deformation ratio (or strain) relationships. Eq.(1) can be rewritten by using elastic stress, $\sigma$, and deformation ratio, $\lambda$, as:

$$\sigma = \left(\frac{\partial u}{\partial \lambda}\right)_{T,V} - T\left(\frac{\partial s}{\partial \lambda}\right)_{T,V} \quad (2)$$

where $u$ and $s$ are internal energy and entropy per a unit volume, respectively. By applying the Maxwell's relations, we obtain

$$\sigma = \left(\frac{\partial u}{\partial \lambda}\right)_{T,V} - T\left(\frac{\partial \sigma}{\partial T}\right)_{\lambda,V} \quad (3).$$

The first and second terms of the right side of Eq.(3) are the energetic and entropic contributions, respectively. The entropic contribution, $-T\left(\frac{\partial \sigma}{\partial T}\right)_{\lambda,V} = \sigma_S$, can be, in principle, obtained by measuring the $\sigma$-$T$ relationship of a material at constant $V$ and $\lambda$. The energetic contribution, $\left(\frac{\partial u}{\partial \lambda}\right)_{T,V} = \sigma_E$, can be obtained by subtracting $\sigma_S$ from $\sigma$ as

$$\sigma_E = \sigma - \sigma_S \quad (4).$$

We synthesized a typical DN gel, a poly(2-acrylamido-2-methylpropanesulfonic acid)/polyacrylamide (PNaAMPS/PAAm) DN gel, as the sample. The first, primary PNaAMPS network is crosslinked with divinylbenzene instead of typical cross-linker $N,N'$-methylene(bis)acrylamide to avoid possible hydrolysis of the cross-linking points and ensure stability of the network during the long experimental process. The PAAm gel was also prepared as the control. The chemical structures and uniaxial tensile test results are shown in Figure 1.

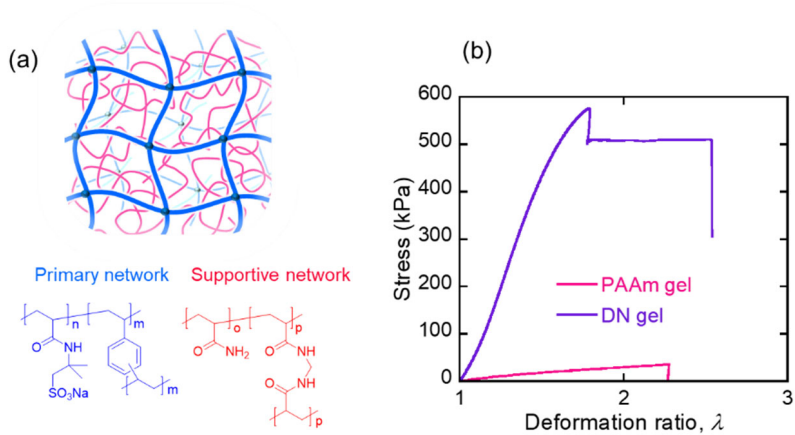

Figure 1. (a) Chemical structure of the double-network gel used in this study. Divinylbenzene, cross-linker for the primary network, is a mixture of isomers. (b) Uniaxial stress-deformation ratio curves of the PAAm gel and DN gel without pre-treatment.

For the separation of the entropic and energetic contributions, the (elastic) stress of the completely relaxed sample at constant $V$ and $\lambda$ under various temperatures is required. To fulfill these requirements as much as possible, the mechanical test was carefully performed as follows. The as-prepared DN gel or the PAAm gel cut into the dumbbell shape (gauge length $L$: 25 mm, width: 7.1 mm) was used as a specimen. Periodic dots with black ink were stamped on a specimen as strain markers (Figure 2(a)) to analyze the accurate deformation ratio of the specimen. The following mechanical experiments on the specimen were performed in the temperature-controlled oil bath (Figure 2(b)). The specimen was soaked in paraffin oil saturated with water to minimize any swelling/de-swelling during the measurement. We used the water-saturated oil because the gel immersed in dried paraffin oil gradually releases water.

The mechanical test was performed following the procedure shown in Figure 2(c). The (engineering) stress of a specimen is defined as the contraction force divided by the initial cross-sectional area of the virgin specimen. The bath temperature was initially set at the lowest target temperature, typically 290 K. The specimen was first uniaxially extended and unloaded. This pre-treatment process aims to cut the short network strands of the primary network prior to the main measurement to avoid the scission of such short strands during the main measurement. [32] The specimen was again extended and kept until reaching the stress plateau, corresponding to the complete relaxation. The specimen was then gradually unloaded while keeping its completely relaxed state at an extremely small velocity of 0.1 mm/min (strain rate of ~$10^{-4}$ s$^{-1}$) until reaching the zero-stress point. The extraction of the pure elastic stress of the sample was confirmed by the overlapping of this unloading curve and the following reloading curve. Thus, the stress obtained by this extremely slow unloading step is deemed as elastic stress. The deformation ratio, $\lambda$, of the specimen during the unloading was determined by the image analysis of the dots stamped on the specimen. The obtained elastic stress-deformation ratio ($\sigma$-$\lambda$) relationship was used in the following analysis. After the unloading test, the bath temperature was raised while the specimen was kept unloaded. After reaching the desired temperature, these extension-relaxation-unloading processes were repeatedly performed at five different temperatures without changing the specimen. Note that we did not apply the pre-treatment and used a faster unloading velocity of 1 mm/min for the PAAm gel because of nearly elastic nature of PAAm hydrogels. [35]

For the thermodynamical analysis with Eq.(4), the mechanical tests should be performed under an isovolumetric condition. However, the real experiment was performed under an isobaric condition. The volume of the specimen during the test, possibly due to swelling/de-swelling or thermal expansion, was continuously tracked by the image analysis of the axial and lateral deformation ratios of the specimen. The volume shrinkage of the specimen through the whole experimental process (~4 days) was found to be less than 2%. Following the previous work, we assume that such a slight change in volume of a gel would not have a critical effect on the results [19].

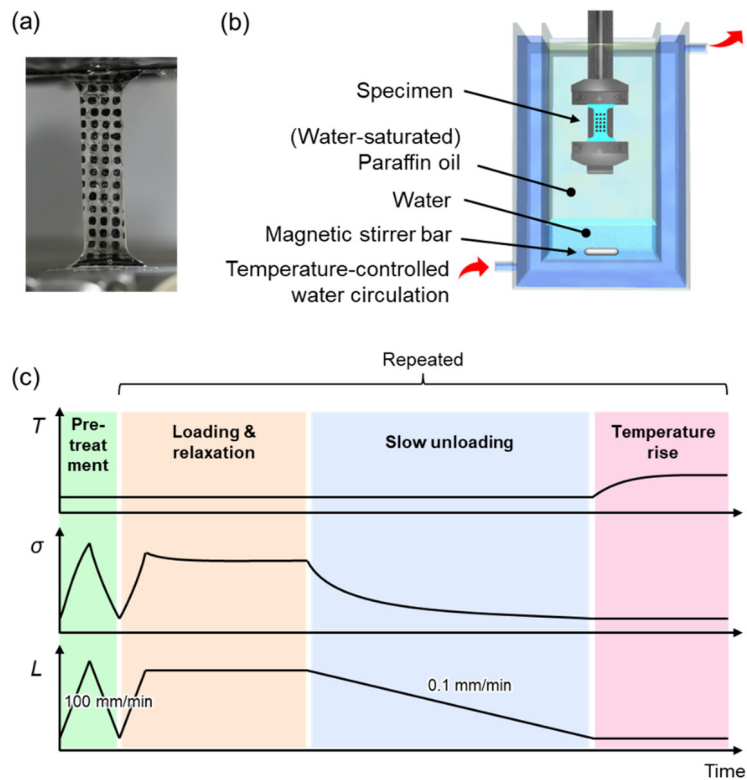

Figure 2. The mechanical testing. (a) A picture of the specimen (double-network gel) with the periodic dots for the strain analysis; (b) the illustration of the setup for the mechanical test. Water-saturated paraffin oil effectively prevents water evaporation from the specimen; (c) The procedure of the mechanical testing, where $T$ is the absolute temperature, $\sigma$ is the elastic stress, and $L$ is the cross-head displacement. Under constant temperature, the sample was extended, kept for complete relaxation, and unloaded very slowly. The procedure was repeated after raising the temperature. Note that the shape of the graphs and the scale of the axes are inaccurate.

Figure 3 shows the elastic stress ($\sigma$)-deformation ratio ($\lambda$) curves of the PAAm and DN gels measured at various temperatures, $T$. The PAAm gel showed a clear positive $\sigma$ dependence on $T$. This is the typical elastic behavior of rubbery materials dominated by entropy. $\sigma$ of the DN gel initially also showed a positive dependence on $T$. However, at large deformation where the remarkable strain hardening occurs, $\sigma$ of the DN gel was almost independent of $T$.

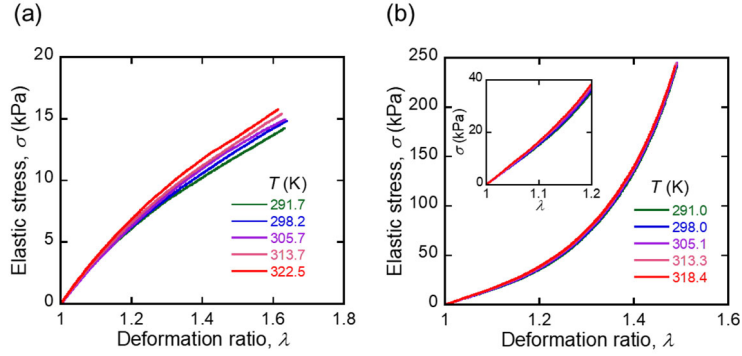

Figure 3. The elastic stress-deformation ratio curves of (a) the PAAm gel and (b) the DN gel measured at various temperatures. The inset in (b) is the enlarged version of the initial regime.

Subsequently, the entropic contribution to the elastic stress at arbitrary deformation ratios was evaluated by Eq.(3). Figure 4(a) shows $\sigma$ of the PAAm gel as a function of $T$ at various $\lambda$. $\sigma$ depends linearly on $T$ in the measured temperature range. We thus drew the linear regression curves for the $\sigma$-$T$ relationships and extrapolated them to 0 K. The reference temperature for the analysis was set at 305 K. The entropic contribution $\sigma_S$ at the reference temperature was estimated from the slopes of the linear regression curves, $\left(\frac{\partial \sigma}{\partial T}\right)_{\lambda,V}$, with Eq.(3). The energetic contribution $\sigma_E = \sigma - \sigma_S$ at the reference temperature was given as the y-intercepts of the regression curves. The slopes of the regression curves increased with $\lambda$, corresponding to an increase in the entropic contribution with deformation. It was also found that the y-intercepts (the energetic contribution) were negative in all measured ranges. These results mean that the elasticity of the PAAm gel consists of a positive entropic contribution and a negative energetic contribution. Moreover, all the regression curves roughly appear to intersect at a particular temperature. The observed negative energetic contribution and the intersection of the regression curves well correspond to the results of the polyethylene glycol hydrogels reported by Sakai *et al*. [21] They proposed the hypothesis that the origin of the negative energetic contributions of the hydrogels is the attractive polymer-water interaction, which would be thus also applicable to the PAAm-water system. The similarity between the PAAm gel data and the polyethylene glycol gel data also suggests that our experimental method is somewhat valid. Figure 4(b) shows the energetic and entropic contributions to the elastic stress of the PAAm gel at various $\lambda$. The entropic contribution is always dominant.

Figure 4(c) shows $\sigma$ dependence on $T$ of the DN gel at various $\lambda$. The linear $\sigma$-$T$ relationships were also found at any $\lambda$. On the other hand, y-intercepts of the linear regression curves are always positive and increase with $\lambda$. These data mean the positive energetic contribution to the elasticity of the highly deformed DN gel. Figure 4(d) shows the energetic and entropic contributions to the elastic stress of the DN gel at various $\lambda$. While the entropic contribution is initially superior, the energetic contribution increases sharply with $\lambda$. At around $\lambda=1.2$ where $\sigma_E$ exceeds $\sigma_S$, the elasticity of the DN gel changes from the entropy-dominated one to the internal energy-dominated one. $\sigma_S$ of the DN gel tends to decrease upon further deformation. This entropic-energetic transition of the elasticity is the same as that of single polymer chains. [14]

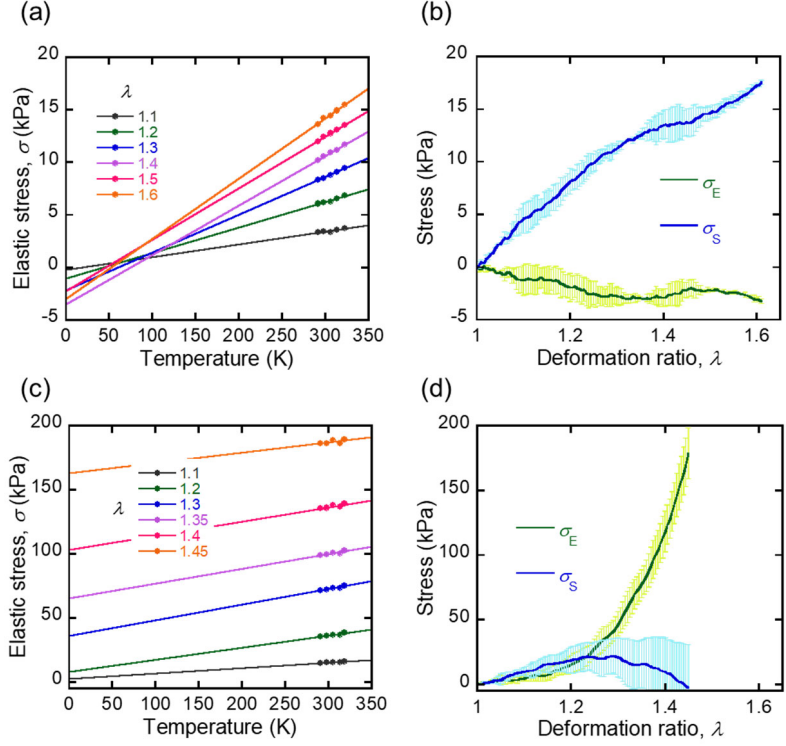

Figure 4. Separation of the energetic and entropic contributions. (a) The elastic stress-temperature relationship of the PAAm gel at various fixed deformation ratios. (b) $\sigma_S$ and $\sigma_E$ of the PAAm gel as a function of $\lambda$ at 305 K. Data are mean ± SD of two independent experiments. (c) The $\sigma$-$T$ relationship of the DN gel at various $\lambda$. (d) $\sigma_S$ and $\sigma_E$ of the DN gel as a function of $\lambda$ at 305 K. Data are mean ± SD of three independent experiments.

**III. A model considering the energetic contribution**

As shown above, the elasticity of the highly deformed DN gels is mainly governed by the positive energetic contribution. Thus, the positive energetic contribution must be taken into account to derive a constitutive mechanical model valid for highly deformed rubbery networks. As such a model, we established the *modified* full-network model. The full-network model hypothesizes a polymer network where end-to-end vector of the strand in the initial state takes any direction with equal probability [36]. We consider the mono-disperse network where all strands have the same number of Kuhn segments $N$ per strand and the same end-to-end distance at an initial state. First, this network is pre-swollen with the swelling ratio $\lambda_s$ along each of three $(x, y, z)$ axes of the network. Next, the pre-swollen network is uniaxially stretched. Strands are assumed to deform affinely to the macroscopic deformation applied to the network. The engineering stress of such a swollen full network at uniaxial elongation ratio $\lambda$ is generally

$$\sigma(\lambda) = \frac{\nu_0}{\lambda_s^3} \frac{R_s}{\lambda} \int_0^{\pi/2} f\left(\frac{R_s \alpha(\theta,\lambda)}{L_c}\right) \frac{\lambda^2 \cos^2\theta - \frac{\sin^2\theta}{2\lambda}}{\alpha(\theta,\lambda)} \sin\theta \, d\theta, \quad (5)$$

where $\nu_0$ is the number density of the strand in the as-prepared network, and $\lambda_s$ is the uniaxial swelling ratio of the swollen network. $L_c = bN$ is a contour length of the strand, where $b$ is the Kuhn length. We assume that the end-to-end distance before applying uniaxial deformation, $R_s$, is given by the root-mean-square end-to-end distance of Gaussian chain, i.e., $R_s = \lambda_s \sqrt{N} b$. Also, $\alpha(\theta,\lambda) \equiv \sqrt{\lambda^2 \cos^2\theta + \frac{\sin^2\theta}{\lambda}}$ is the stretch ratio $(R/R_s)$ of strands at $\lambda$ that had polar angle $\theta$ measured between the end-to-end vector and elongation axis before applying uniaxial deformation, where $R$ is the end-to-end distance of the strand of interest. In the original full-network model, the FJC model equation is used as the force function in Eq.(5). [36] On the other hand, we adopt the m-FJC model given by Eq.(6) as the force function to construct the *modified* full-network model,

$$\frac{R}{L_c} = \left(\coth\left(\frac{bf}{k_B T}\right) - \frac{k_B T}{bf}\right)\left(1 + \frac{f}{f_s}\right), \quad (6)$$

where $f_s$ is the characteristic force representing energetic elasticity of the strand [11]. To implement the force-extension relation given by Eq.(6) into Eq.(5), we derived an approximate expression of the force of the m-FJC chain as a function of the extension as

$$f^*(r) \simeq \frac{1}{2}\sqrt{(f_s^* r)^2 - 2(f_s^* - 1)f_s^* r + (f_s^* + 1)^2} - \left(\frac{9}{f_s^*} + \left(\frac{f_s^*}{1 + f_s^*}\right)^3\right) r^2$$
$$+ \frac{(f_s^* + 2)(f_s^* + 3)}{2(f_s^* + 1)} r - \frac{f_s^* + 1}{2}, \quad (7)$$

where $f^* \equiv bf/k_B T$, $f_s^* \equiv bf_s/k_B T$, and $r \equiv R/L$ are nondimensionalized quantities. Relative error $|(f \text{ from Eq.(7)}) - (f \text{ from Eq.(6)})| / (f \text{ from Eq.(6)})$ is not more than 3% for typical value of $f_s^*$ ($= 10^3 \sim 10^4$). The stress of the modified full-network model was numerically calculated by substituting Eq.(7) into Eq.(5). We set $N$, $\nu_0$, and $f_s$ as the fitting parameters, whereas the other variables were determined based on the experiments and literature (see Appendix A) [7,37]. We also used the original full-network model for the comparison (see Appendix B).

As shown in Figure 5(a), the $\sigma$-$\lambda$ curve of the PAAm gel is well fitted both by the original and modified full-network models because the deformation of the PAAm network is so small compared with its theoretical stretching limit that the energetic contribution does not appear. On the other hand, the $\sigma$-$\lambda$ curve of the DN gel was adequately fitted by the modified full-network model with reasonable fitting parameters but poorly fitted with the original full-network model with any fitting parameters (Figure 5(b)). Moreover, as shown in Figure 5(c-f), the modified full-network model also reproduces the temperature dependence of the σ-λ curve of the PAAm gel and DN gel. These data strongly suggest the necessity of considering the positive energetic contribution to describe the elastic response of highly deformed rubbery materials including DN gels. Note that because the mechanical properties of DN gels before yielding are dominated by the rigid primary network, we used a model with single network parameters (representing the primary network) to reproduce the data of the DN gels with two networks.

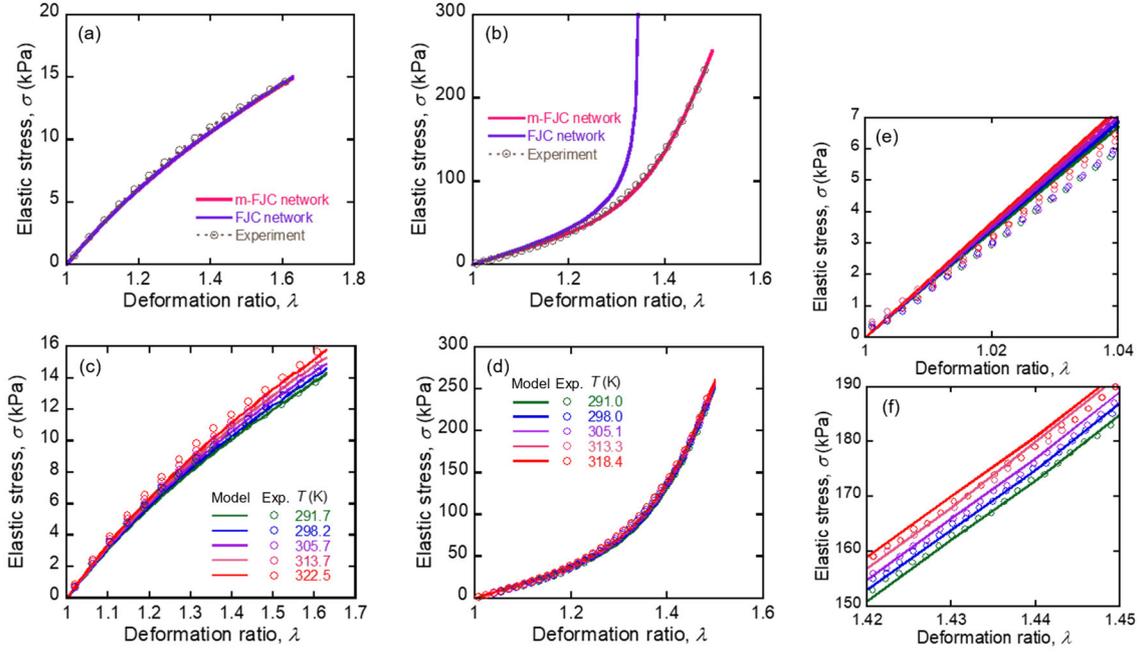

Figure 5. The best fitting curves for the σ-λ behaviors of the (a) PAAm gel at 305.7 K and (b) DN gel at 305.1 K with m-FJC-based and FJC-based full-network models. (c-f) The σ-λ behaviors of the (c) PAAm gel and (d-f) DN gel at various temperatures fitted by the m-FJC-based model; (d) whole data, (e) at small deformation dominated by the entropic elasticity, and (f) at large deformation dominated by the energetic elasticity. The same fitting parameters were used for a series of the data of the same gel at different temperatures.

**IV. Conclusions**

We experimentally found that the elasticity of the highly extended DN gel is dominated by the positive energetic contribution. As far as the authors know, this is the first experimental report of remarkably positive energetic elasticity of the ultimately deformed polymer network possibly due to the chemical bond deformation. This research suggests the importance of linking the elasticity of single polymer chains and rubbery networks. In the mechanics of single polymer chains, chemistry-based theoretical studies, e.g., quantum mechanical calculations or models considering activation energies, have been performed to discuss the chemical origin of the energetic elasticity in the ultimately stretched regime. [8,9,38] Similarly, to discuss the mechanics of ultimately deformed rubbery materials, it is necessary to develop elasticity models considering chemistry-based findings of the single chain mechanics. This research also shows that the rubbery materials having the double-network structure (DN materials) are good candidates for research on ultimately deformed rubbery networks, following our previous research that first reveals the inversion of the mechanical-swelling coupling of an ultimately deformed gel. [27] A drawback of this study is the quantifiability issue in the separation of the energetic and entropic contributions due to ignoring the slight volume changes of the specimen. To overcome this issue, the DN gels with a non-volatile solvent, which allows the isobaric mechanical test without solvent migration, are anticipated. It is also essential to develop a method to correct the isobaric test results of ultimately deformed rubbery materials to the isovolumetric test results. Such correction formulas have been proposed, but they are available only for the Gaussian polymer network. [39] It would be necessary to find the constitutive equation for the rubbery network that undergoes ultimate deformation and to derive a correction formula applicable to a rubbery material following this constitutive equation.

## V. Methods

**Preparation of the double-network hydrogel and the polyacrylamide hydrogel**

PNaAMPS/PAAm double-network (DN) gels were synthesized via two-step sequential free-radical polymerization. Initially, a 20 ml solution of the first network precursor was prepared by mixing 1.4 M 2-acrylamido-2-methylpropanesulfonic acid sodium salt (NaAMPS, Toa Gosei Co., Ltd., Japan) as the monomer, 56 mM divinylbenzene (DVB, FUJIFILM Wako Pure Chemical Corporation, Inc., Japan) as the cross-linker, 14 mM 2-oxoglutaric acid (FUJIFILM Wako Pure Chemical Corporation, Inc., Japan) as the initiator, 9 ml *N*,*N*-dimethylformamide (FUJIFILM Wako Pure Chemical Corporation, Inc., Japan), and pure water. The solution was poured into molds formed by two 3 mm-thick glass plates separated by a 0.5 mm-thick U-shaped silicone rubber spacer. The molds were irradiated with 365 nm intense UV light (340 mW/cm$^2$) for 30 minutes to prepare the PNaAMPS gels. The obtained PNaAMPS gels were immersed in an excess second network precursor aqueous solution containing 2 M acrylamide (AAm, Junsei Chemical Co., Ltd., Japan, recrystallized from acetone) as the monomer, $4 \times 10^{-4}$ M *N*,*N'*-methylenebisacrylamide (MBAA, FUJIFILM Wako Pure Chemical Corporation, Inc., Japan) as the cross-linker, and $4 \times 10^{-4}$ M 2-oxoglutaric acid as the initiator for 1 day. The swollen gels were then sandwiched between two glass plates and transferred to an argon blanket. DN gels were prepared by irradiation with 365 nm UV light (4 mW/cm$^2$) for 8 hours to synthesize the PAAm network in the presence of the PNaAMPS network. The as-prepared DN gels were directly used for the tests.

Polyacrylamide gels were synthesized as follows. The gel precursor aqueous solution was prepared by mixing 2 M AAm, 10 mM MBAA, and 2 mM 2-oxoglutaric acid with pure water. The solution was poured into molds formed by two soda-lime glass plates separated by a 1 mm-thick U-shaped silicone rubber spacer. PAAm gels were prepared by irradiation with 365 nm UV light (4 mW/cm$^2$) for 8 hours. The as-prepared PAAm gels were directly used for the tests.

**Uniaxial tensile test**

The as-prepared gels were cut into the dumbbell-shaped specimen standardized as JIS-K6251-7 (gauge length: 12 mm, width: 2 mm) with a press cutter. The uniaxial tensile test of a specimen was performed by the mechanical tester equipped with a non-contact video extensometer (Instron 5965 with AVE, Instron Co., USA). The two white dots were stamped on the sample as strain markers. The (engineering) stress was defined as the contraction force divided by the initial cross-sectional area of the specimen, and the deformation ratio was defined as the distance between the two dots of a specimen divided by that at the relaxed state. The tensile velocity was 100 mm/min.

**Super-slow uniaxial unloading test in oil**

The as-prepared gels were cut into the dumbbell-shaped specimen with a length of 25 mm and a width of 7.1 mm by the laser cutter (PLS4.75, Universal Laser Systems Inc., USA). Periodic black dots were stamped on the cut specimens as a strain marker. The test was performed with the mechanical tester (TENSILON RTG-1310, Orientec Co., Japan) equipped with a transparent, water-jacketed oil bath filled with water-saturated paraffin oil (Figure 1). The temperature of the oil was controlled by circulating temperature-controlled water into a jacket and logged with the temperature logger (TR-71U, T&D Co., Japan). A specimen set to the tester was always completely in the paraffin oil during the test. Photos of a specimen were frequently taken by a camera during the test for a deformation analysis. The (lateral) deformation ratio $\lambda$ and transversal deformation ratio $\lambda_t$ were obtained by the image analysis. Specifically, $\lambda$ and $\lambda_t$ of the specimen were determined by tracking the distance between two horizontally adjacent dots and between two laterally adjacent dots stamped on the specimen, respectively. The normalized volume of the specimen was calculated by $\lambda\lambda_t^2$ under the assumption of isotropic swelling.

The temperature of the oil (and specimen) was initially set at the lowest target temperature, typically 290 K. Only for the DN gel, a specimen was pre-treated before the main experiment, *i.e.* a DN gel specimen was uniaxially extended 20 mm and immediately unloaded to the zero-stress point with the tensile velocity of 100 mm/min. The pre-treated DN specimen or virgin PAAm specimen was uniaxially extended 17.5 mm at 100 mm/min. After reaching the desired length, the specimen was kept until reaching the stress plateau (typically around 4 h for a DN specimen). Then, the specimen was gradually unloaded at 0.1 mm/min until reaching zero-stress point. The stress obtained from this process is called elastic stress. The obtained elastic stress-lateral deformation ratio ($\sigma$-$\lambda$) relationship was used in the analysis. After the complete unloading, the bath temperature was raised to the next target temperature without removing the specimen. After reaching the target temperature, the specimen was again uniaxially extended 17.5 mm at 100 mm/min, kept until reaching the stress plateau, and gradually unloaded at 0.1 mm/min until reaching zero-stress point. These processes were repeated at various target temperatures.

**Acknowledgments**

This work was supported by JST PRESTO (Grant No. JPMJPR2098), JST FOREST (Grant No. JPMJFR221X) and JSPS KAKENHI (Grant Nos. 22H04968, 22K21342, 24H00848). The Institute for Chemical Reaction Design and Discovery (WPI-ICReDD) was established by the World Premier International Research Initiative (WPI), MEXT, Japan. The authors thank Dr. Yoshinori Katsuyama (Hokkaido University) for his extensive support on the experimental setup. T. N. thanks Dr. Naomichi Sakumichi (The University of Tokyo) for the fruitful discussion.

**Appendices**

**APPENDIX A: Determination of the parameters in the model**

To apply the network model, several parameters need to be determined. Swelling ratio, $\lambda_s$, was set to 1.0 for the (as-prepared) PAAm gel and 2.10 for the DN gel from the experimental data. For the DN gel, $\lambda_s$ was determined as the thickness of the DN gel divided by that of the as-prepared first network gel. Kuhn length, $b$, was set to 0.6 nm for the PAAm and 0.5 nm for the DN gel based on the literature values [7,37]. The experimentally observed temperature was used as $T$. Other variables, $\nu_0$, $N$, and $f_s$, were set as the fitting parameters.

For the DN gel, the best fitting parameters at $T \approx 305$ K are $\nu_0$=21.3 mol/m$^3$, $N$=8, and $f_s$=4.3 nN. The experimentally estimated $\nu_0$ of the primary PAMPS network is 30 mol/m$^3$ based on the data of our previous literature [S4]. The best fitting parameter $\nu_0$=21.3 mol/m$^3$ is enough close to this experimental value. For $N$=8, since 4 mol% of the two-functional cross-linker was used relative to the monomer AMPS for preparation of the primary network, the ideal number of monomer units in the strand is 12.5. Since 1 Kuhn segment of PAMPS (0.5 nm) contains approximately 2 monomer units, the ideal $N$ is 6.5, which is enough close to the best fitting parameter. Finally, according to the literature, $f_s$ of PAMPS in pure water is 55 nM, which is one order larger than the best fitting parameter $f_s$=4.3 nN [37]. However, the literature also shows that $f_s$ of PAMPS significantly decreases with increase of the counterion concentration of the environment. Since inside of the DN gel can be considered as polyelectrolyte solution due to the PAMPS primary network, $f_s$ of PAMPS in the DN gel would be remarkably lower than that in pure water. Thus, we think it is acceptable to say that the best fitting parameter $f_s$=4.3 nN is reasonable.

For the PAAm gel, the best fitting parameters at $T \approx 305$ K are $\nu_0$=4.9 mol/m$^3$, $N$=100, and $f_s$=17.5 nN. Note that the fitting curves for the PAAm gel at the given condition is not sensitive to $N$ and $f_s$ at $\lambda$<1.63 because effect of the energetic elasticity does not appear, thus $N$ and $f_s$ for the PAAm gel can take a wide range of values. $\nu_0$=4.9 mol/m$^3$ for the PAAm gel is consistent with the experimentally estimated value. Young's modulus, $E$, of the PAAm gel at $T \approx 305$ K was 36 kPa obtained from the initial slope of the $\sigma$-$\lambda$ curve (Figure 1). Using the classical rubber elasticity theory for an Affine network, $E = 3\nu_0 RT$, we estimated $\nu_0$ as 4.7 mol/m$^3$ based on the modulus.

**APPENDIX B: The (original) full-network model**

The original, non-modified full-network model was based on the force function of the FJC model,

$$\frac{R}{L_c} = \left(\coth\left(\frac{bf}{k_B T}\right) - \frac{k_B T}{bf}\right), \qquad (8)$$

where the corresponding Cohen-Padé approximation is, [40]

$$f^* \simeq \frac{r(3-r^2)}{1-r^2}. \qquad (9)$$

The relative error between the forces calculated by Eqs.(8) and (9) is typically less than 5%. The full-network model was constructed by substitution of Eq.(9) into Eq.(5).

**References**

[1] H. M. James and E. Guth, *Theory of the Elastic Properties of Rubber*, J. Chem. Phys. **11**, 455 (1943).

[2] P. J. Flory, *Network Structure and the Elastic Properties of Vulcanized Rubber.*, Chem. Rev. **35**, 51 (1944).

[3] W. Kuhn and F. Grün, *Beziehungen zwischen elastischen Konstanten und Dehnungsdoppelbrechung hochelastischer Stoffe*, Kolloid-Z. **101**, 248 (1942).

[4] L. R. G. Treloar, *The Elasticity of a Network of Long-Chain Molecules—II*, Trans Faraday Soc **39**, 241 (1943).

[5] E. M. Arruda and M. C. Boyce, *A Three-Dimensional Constitutive Model for the Large Stretch Behavior of Rubber Elastic Materials*, J. Mech. Phys. Solids **41**, 389 (1993).

[6] H. Li, B. Liu, X. Zhang, C. Gao, J. Shen, and G. Zou, *Single-Molecule Force Spectroscopy on Poly(Acrylic Acid) by AFM*, Langmuir **15**, 2120 (1999).

[7] W. Zhang, S. Zou, C. Wang, and X. Zhang, *Single Polymer Chain Elongation of Poly( N - Isopropylacrylamide) and Poly(Acrylamide) by Atomic Force Microscopy*, J. Phys. Chem. B **104**, 10258 (2000).

[8] P. E. Marszalek and Y. F. Dufrêne, *Stretching Single Polysaccharides and Proteins Using Atomic Force Microscopy*, Chem. Soc. Rev. **41**, 3523 (2012).

[9] F. Oesterhelt, M. Rief, and H. E. Gaub, *Single Molecule Force Spectroscopy by AFM Indicates Helical Structure of Poly ( Ethylene-Glycol ) in Water*, New J. Phys. **1**, 1 (1999).

[10] Y. Bao and S. Cui, *Single-Chain Inherent Elasticity of Macromolecules: From Concept to Applications*, Langmuir **39**, 3527 (2023).

[11] S. B. Smith, Y. Cui, and C. Bustamante, *Overstretching B-DNA: The Elastic Response of Individual Double-Stranded and Single-Stranded DNA Molecules*, Science **271**, 795 (1996).

[12] A. Kolberg, C. Wenzel, K. Hackenstrass, R. Schwarzl, C. Rüttiger, T. Hugel, M. Gallei, R. R. Netz, and B. N. Balzer, *Opposing Temperature Dependence of the Stretching Response of Single PEG and PNiPAM Polymers*, J. Am. Chem. Soc. **141**, 11603 (2019).

[13] S. Lu, W. Cai, S. Zhang, and S. Cui, *Single-Chain Mechanics at Low Temperatures: A Study of Three Kinds of Polymers from 153 to 300 K*, Macromolecules **56**, 3204 (2023).

[14] S. Wang, S. Panyukov, M. Rubinstein, and S. L. Craig, *Quantitative Adjustment to the Molecular Energy Parameter in the Lake–Thomas Theory of Polymer Fracture Energy*, Macromolecules **52**, 2772 (2019).

[15] M. Shen and M. Croucher, *Contribution of Internal Energy to the Elasticity of Rubberlike*

*Materials*, J. Macromol. Sci. Part C Polym. Rev. **12**, 287 (1975).

[16] R. L. Anthony, R. H. Caston, and E. Guth, *Equations of State for Natural and Synthetic Rubber-like Materials. I. Unaccelerated Natural Soft Rubber*, J. Phys. Chem. **46**, 826 (1942).

[17] G. Allen, M. J. Kirkham, J. Padget, and C. Price, *Thermodynamics of Rubber Elasticity at Constant Volume*, Trans. Faraday Soc. **67**, 1278 (1971).

[18] T. Aoyama and K. Urayama, *Negative and Positive Energetic Elasticity of Polydimethylsiloxane Gels*, ACS Macro Lett. **12**, 356 (2023).

[19] A. J. Licup, A. Sharma, and F. C. MacKintosh, *Elastic Regimes of Subisostatic Athermal Fiber Networks*, Phys. Rev. E **93**, 012407 (2016).

[20] J.-B. Le Cam, *Strain-Induced Crystallization in Rubber: A New Measurement Technique*, Strain **54**, e12256 (2018).

[21] Y. Yoshikawa, N. Sakumichi, U. Chung, and T. Sakai, *Negative Energy Elasticity in a Rubberlike Gel*, Phys. Rev. X **11**, 011045 (2021).

[22] N. Sakumichi, Y. Yoshikawa, and T. Sakai, *Linear Elasticity of Polymer Gels in Terms of Negative Energy Elasticity*, Polym. J. **53**, 1293 (2021).

[23] A. N. Gent and W. V. Mars, *Chapter 10 - Strength of Elastomers*, in *The Science and Technology of Rubber (Fourth Edition)*, edited by J. E. Mark, B. Erman, and C. M. Roland (Academic Press, Boston, 2013), pp. 473–516.

[24] M. C. Boyce and E. M. Arruda, *Constitutive Models of Rubber Elasticity: A Review*, Rubber Chem. Technol. **73**, 504 (2000).

[25] J. P. Gong, Y. Katsuyama, T. Kurokawa, and Y. Osada, *Double-Network Hydrogels with Extremely High Mechanical Strength*, Adv. Mater. **15**, 1155 (2003).

[26] J. P. Gong, *Why Are Double Network Hydrogels so Tough?*, Soft Matter **6**, 2583 (2010).

[27] Y. Zhang, K. Fukao, T. Matsuda, T. Nakajima, K. Tsunoda, T. Kurokawa, and J. P. Gong, *Unique Crack Propagation of Double Network Hydrogels under High Stretch*, Extreme Mech. Lett. **51**, 101588 (2022).

[28] T. Matsuda, T. Nakajima, Y. Fukuda, W. Hong, T. Sakai, T. Kurokawa, U. I. Chung, and J. P. Gong, *Yielding Criteria of Double Network Hydrogels*, Macromolecules **49**, 1865 (2016).

[29] R. E. Webber, C. Creton, H. R. Brown, and J. P. Gong, *Large Strain Hysteresis and Mullins Effect of Tough Double-Network Hydrogels*, Macromolecules **40**, 2919 (2007).

[30] T. Nakajima, T. Kurokawa, S. Ahmed, W. L. Wu, and J. P. Gong, *Characterization of Internal Fracture Process of Double Network Hydrogels under Uniaxial Elongation*, Soft Matter **9**, 1955 (2013).

[31] Y. H. Na, Y. Tanaka, Y. Kawauchi, H. Furukawa, T. Sumiyoshi, J. P. Gong, and Y. Osada, *Necking Phenomenon of Double-Network Gels*, Macromolecules **39**, 4641 (2006).

[32] C. Imaoka, T. Nakajima, T. Indei, M. Iwata, W. Hong, A. Marcellan, and J. P. Gong, *Inverse*

*Mechanical-Swelling Coupling of a Highly Deformed Double-Network Gel*, Sci. Adv. **9**, eabp8351 (2023).

[33] T. Matsuda, R. Kawakami, R. Namba, T. Nakajima, and J. P. J. P. Gong, *Mechanoresponsive Self-Growing Hydrogels Inspired by Muscle Training*, Science **363**, 504 (2019).

[34] T. Nakajima, T. Kurokawa, H. Furukawa, and J. P. J. P. Gong, *Effect of the Constituent Networks of Double-Network Gels on Their Mechanical Properties and Energy Dissipation Process*, Soft Matter **16**, 8618 (2020).

[35] M. Fujine, T. Takigawa, and K. Urayama, *Strain-Driven Swelling and Accompanying Stress Reduction in Polymer Gels under Biaxial Stretching*, Macromolecules **48**, 3622 (2015).

[36] P. D. Wu and E. Van Der Giessen, *On Improved 3-D Non-Gaussian Network Models for Rubber Elasticity*, Mech. Res. Commun. **19**, 427 (1992).

[37] S. Cui, C. Liu, Z. Wang, X. Zhang, S. Strandman, and H. Tenhu, *Single Molecule Force Spectroscopy on Polyelectrolytes: Effect of Spacer on Adhesion Force and Linear Charge Density on Rigidity*, Macromolecules **37**, 946 (2004).

[38] S. Cui, Y. Yu, and Z. Lin, *Modeling Single Chain Elasticity of Single-Stranded DNA: A Comparison of Three Models*, Polymer **50**, 930 (2009).

[39] W. W. Graessley and L. J. Fetters, *Thermoelasticity of Polymer Networks*, Macromolecules **34**, 7147 (2001).

[40] A. Cohen, *A Padé Approximant to the Inverse Langevin Function*, Rheol. Acta **30**, 270 (1991).